\documentclass[letter]{aa} 
\usepackage{graphicx}
\begin{document}
   \title{Detection of nonthermal emission from the bow shock of a 
massive runaway star}

   \author{P. Benaglia
          \inst{1,2}\fnmsep\thanks{Member of CONICET},
          G. E. Romero
          \inst{1,2,\star},
          J. Mart\'{\i}
          \inst{3},
          C. S. Peri
          \inst{1}\fnmsep\thanks{Fellow of CONICET}        
          \and
          A. T. Araudo\inst{1,2,\star\star}
          }

   \institute{Instituto Argentino de Radioastronom\'{\i}a, 
CCT-La Plata (CONICET), C.C.5, (1894) Villa Elisa, Argentina\\
   \email{[pbenaglia;cperi]@fcaglp.unlp.edu.ar, 
[romero;anabella]@iar-conicet.gov.ar}
   \and 
    Facultad de Cs. Astron\'omicas y Geof\'{\i}sicas, UNLP, 
Paseo del Bosque s/n, (1900) La Plata, Argentina
   \and 
    Departamento de F\'{\i}sica (EPS), Universidad de Ja\'en,
Campus Las Lagunillas s/n, Edif. A3, 23071 Ja\'en, Spain\\              
   \email{jmarti@ujaen.es}
             }

\authorrunning{Benaglia et~al.}
\titlerunning{Nonthermal emission from the bow shock of a runaway star}

   \date{Received; accepted }

  \abstract 
{The environs of massive, early-type stars have been 
  inspected in recent years in the search for sites where particles can be 
  accelerated up to relativistic energies. Wind regions 
  of massive binaries that collide have already been established as sources of 
  high-energy emission; however, there is a different 
  scenario for massive stars where strong shocks can also be 
  produced: the bow-shaped region of matter piled up by the action of
  the stellar strong wind of a runaway star interacting with the 
  interstellar medium.} 
{We study the bow-shock region produced by a very massive runaway star,
  BD+43$^\circ$3654, to look for nonthermal radio emission as evidence of a
  relativistic particle population.} 
{We observed the field of  BD+43$^\circ$3654 at two frequencies, 1.42 and 
  4.86 GHz, with the Very Large Array (VLA), and obtained a spectral index map 
  of the radio emission.} 
{We have detected, for the first time, nonthermal radio emission from the 
  bow shock of a massive runaway star.} 
{After analyzing  the radiative mechanisms that can be 
  at work, we conclude that the region under study could produce enough
  relativistic particles whose radiation might be detectable by 
  forthcoming gamma-ray instruments, like CTA North.}


   \maketitle
%

\section{Introduction}

Early-type stars with high peculiar velocities (i.e. runaway stars, 
with velocities 
$v_* > 30$ km s$^{-1}$, e.g.  Gies \& Bolton 1986) are uncommon. For 
instance, Ma\'{\i}z-Apell\'aniz et al. (2004, Galactic O star catalog 
for $V < 8$ stars) lists $\sim$ 8\% runaway stars out of 
370. These particular stars can be identified by the perturbation they 
produce in the ambient medium (e.g. Kobulnicky et al. 2010 and references 
therein). When the strong winds of runaway OB stars sweep relatively 
large amounts of gas and dust, the material piles up in the so-called 
stellar bow shock. Bow shocks develop as arc-shaped structures, with 
bows pointing in the same direction as the stellar velocity, while the 
star moves supersonically in the surrounding interstellar medium (ISM). The 
winds are confined by the ram pressure of the ISM, at distances from 
the star determined by momentum balance. The stellar and 
shock-excited radiation heats the dust and gas swept by the bow shock. 
The dust, in turn, re-radiates the energy as mid-to-far IR excess flux.

As soon as IRAS images became available, Van Buren \& McCray (1988) 
looked for bow-shaped features near high-velocity O stars (see also Van 
Buren et al. 1995 and Noriega-Crespo et al. 1997). The authors detected 
an IR candidate close to the O supergiant BD+43$^{\circ}3654$ 
($\alpha,\delta$[J2000] = $20^{\rm h}33^{\rm m}36.077^{\rm s}, +43^{\circ} 
59' 07.40''$; $l, b = 82.41^\circ, +2.33^\circ$).

Recently, Comer\'on \& Pasquali (2007) related the star 
BD+43$^{\circ}3654$ to a 
bow shock detected with the Midcourse Space eXperiment (MSX) at D and E 
bands. They studied the stellar motion relative to the surrounding 
material and proposed the star is a runaway member 
from Cyg OB2 association. Comer\'on and Pasquali determined a spectral 
type O4 If, and derived an age of about 1.6 Myr and a stellar mass of 
$\sim$ 70 M$_\odot$, which makes the star one of the three more massive 
runaway stars known so far. On the basis of these estimates, the 
authors favor a dynamical ejection scenario (see Hoogerwerf et al. 
2000, 2001 for reviews about the origin of runaway stars).

Gravamazde \& Bomans (2008) suggest instead that 
BD+43$^{\circ}\,3654$ is part of a stellar system, formed by a close 
encounter between two tight massive binaries in the core of Cyg OB2. 
The star should be a blue straggler to match the timescales 
involved in their hypothesis.  Kobulnicky et al. (2010) measured a 
heliocentric radial velocity of $-66.2\pm9.4$ km
s$^{-1}$, and derived a stellar mass-loss rate of $1.6 \times 10^{-4}$ 
M$_\odot$ yr$^{-1}$.

We analyzed data from the NRAO-VLA Sky Survey (NVSS, Condon et al. 
1998). The images revealed a coma-shaped source of $\sim$ 7 arcmin, 
spatially coincident with the MSX structure (see Figure 1; NVSS angular 
resolution: 45''; rms noise: 1 mJy beam$^{-1}$). No point sources above 
5$\sigma$ (40 mJy) that are positionally coincident with the MSX source are 
detected in the MIT-Green Bank Survey (GB6, Griffith et al. 1991). 
Inspection of the continuum emission at 408 and 1420 MHz with the 
CGPS Survey (Taylor et al. 2003, angular resolutions of $3.4'$ and $1'$) 
confirms that the region centered on Cyg OB2 is complex and has 
strong emission on various angular scales (Peri et al. 2010).

\begin{figure}
\begin{center}
\includegraphics[bb= 157 154 470 720,
width=4.8cm,angle=-90]{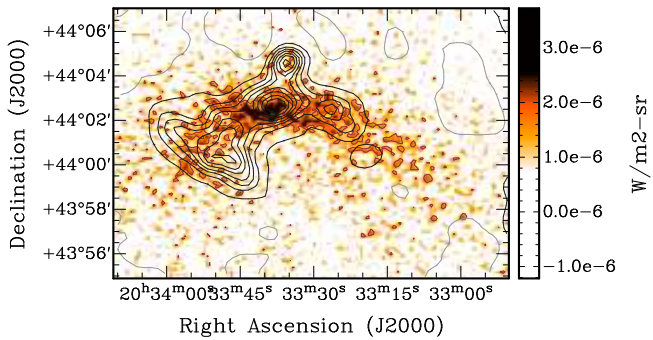}
\caption{MSX-D band image (color scale) superposed to 1.4 GHz-NVSS 
contours. Levels are: $-$2, 2 (2$\sigma$), 5, 8, 11, 15, 19, 24, 29, 50, 70,
 and 90 mJy beam$^{-1}$. 
}
\label{nvss-msx}
\end{center}
\end{figure}

A radio study of the bow shock can shed light on the physical processes 
that give rise to high-energy emission from a stellar source, 
regardless of the history of the runaway star. The shock can accelerate 
particles up to relativistic energies by Fermi mechanism. Energetic 
electrons will cool through synchrotron radiation, producing a 
nonthermal radio source. We carried out radio observations at two 
frequencies to study the nature of the emission from the bow shock of 
BD+43$^{\circ}\,3654$.

In this {\sl Letter} we present the results of the radio observations 
in the form of a 
spectral index map of the bow-shock region of the runaway star and 
distinguish between 
thermal and nonthermal emission regions. We briefly discuss the 
issue of whether a bow shock 
could produce high-energy emission enough to be detected with 
instruments like Fermi or the 
future Cherenkov Telescope Array (CTA) and the conditions that 
must be fulfilled to achieve detection.
 

\section{Observations and results}

We carried out continuum observations with the Very Large Array 
(NRAO\footnote{The National Radio Astronomy Observatory is a facility of 
the National Science Foundation operated under cooperative agreement by 
Associated Universities, Inc.}) in two array configurations: C 
at 1.42 GHz in April 2008 and D at 4.86 GHz in August 2008. The set up 
allowed mapping largest structures of the size of the MSX source and 
also 
guaranteed good matching beams of $12''$. The flux calibrator used was 
0137+331\footnote{Calibrator details can be found at www.vla.nrao.edu/.}. 
Phase calibrator scans --2052+365 at L band and 2007+404 at C band-- were 
interleaved with target scans. The total time on source was 3h at each 
band and 50 MHz for  the total bandwidth.

%
%

The data was calibrated with the AIPS package in the standard way and 
analyzed with the Miriad routines. We used {\sl imagr} to produce 
robust-weighted images. Figure 2 presents the resulting images after 
primary beam correction, re-gridded with the same synthesized beam. At 
both frequencies there is emission along the extension of the MSX source. 
The radio source is larger at 1.42 GHz, toward the eastern half of IR 
contours of the bow shock (increasing right ascensions). Clearly, the 
spectral index $\alpha$ ($S_{\nu} \propto \nu^{\alpha}$) is negative in 
there. In both images, the rms attained is similar (0.3 mJy beam$^{-1}$ 
at 1.42 GHz and 0.2 at 4.8, $12''$ synthesized beam).

A detached ellipsoidal source at $\alpha,\delta$[J2000] = $20^{\rm h}$ 
$33^{\rm m}35^{\rm s}, 44^{\circ} 04' 30''$, called ES here, is the strongest 
one in the field. The fluxes are $S_{\rm ES} (\rm 1.4GHz)= 105\pm10$ mJy 
and $S_{\rm ES}(\rm 4.8GHz)= 95\pm 5$ mJy. ES spectral index $<\alpha> = 
-0.1\pm0.1$ is characteristic of an optically thin H{\sc ii} region.

\begin{figure}
\begin{center}
\includegraphics[width=6cm,angle=-90]{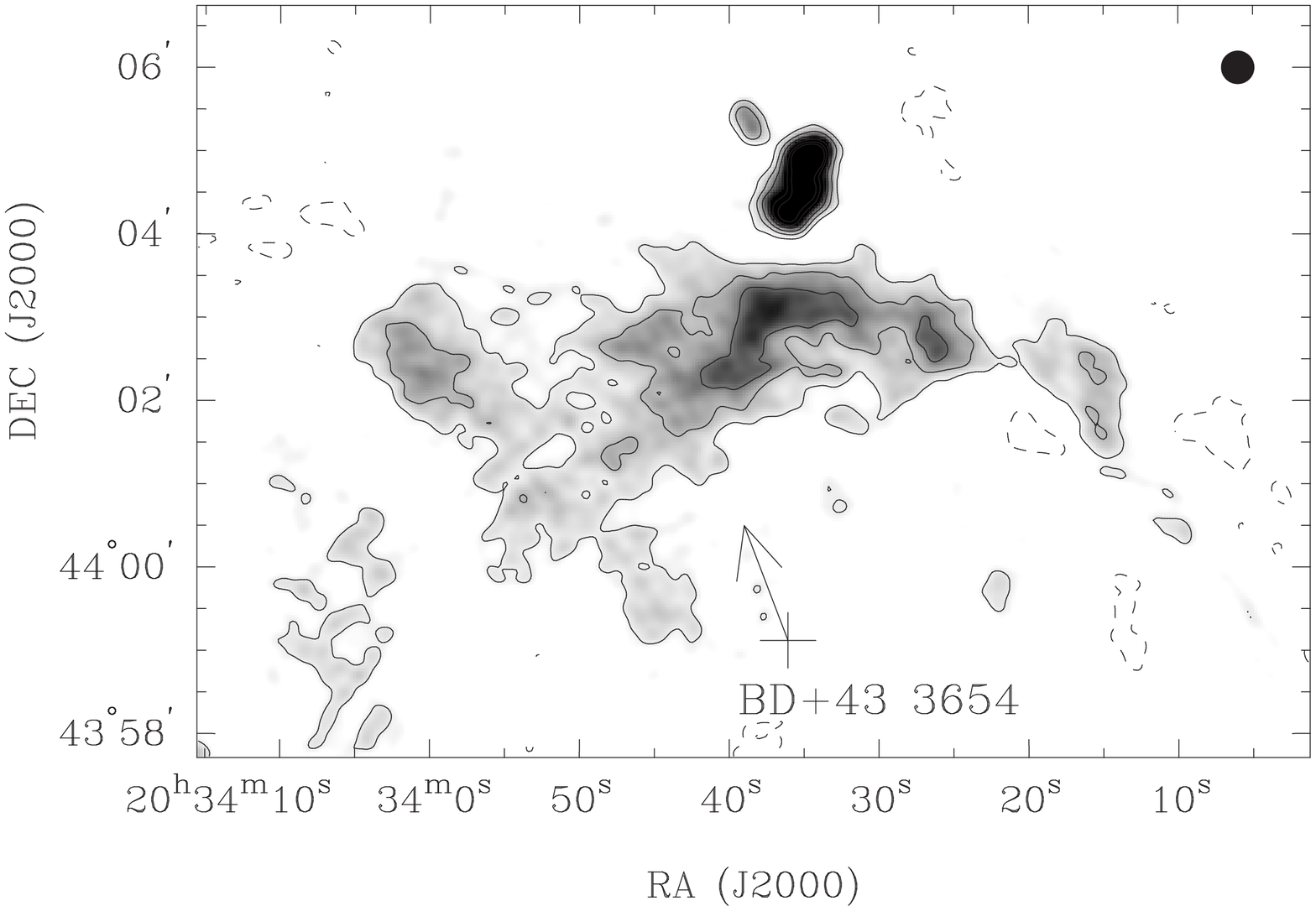}
\includegraphics[width=6cm,angle=-90]{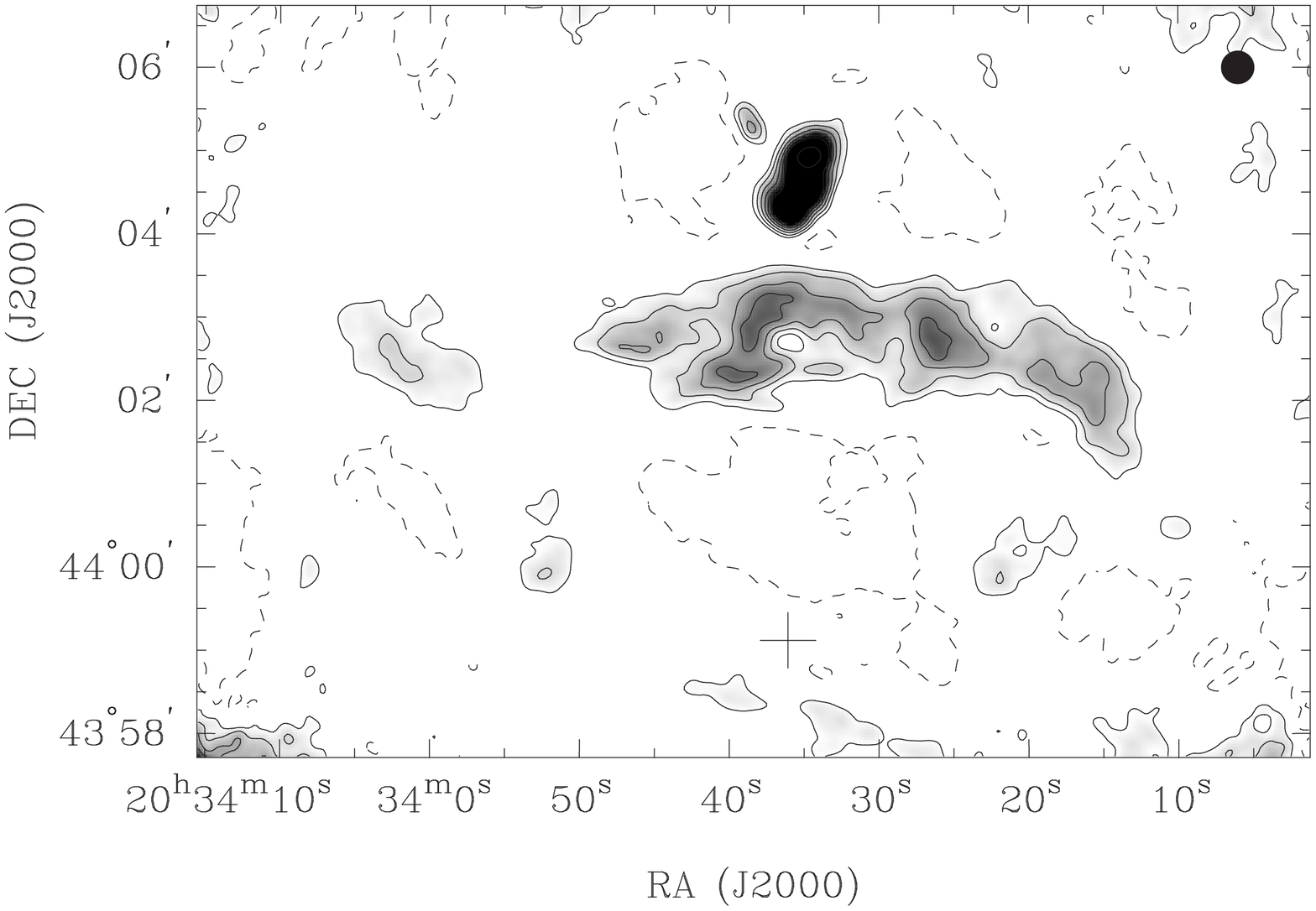}
\caption{Continuum emission at 1.42 GHz (upper panel), and
at 4.86 GHz (lower panel). Contour levels are $-3$, 3, 
6, 10, 15, 20, 25, and 60 times the rms of 0.3 and 0.2 mJy beam$^{-1}$. 
BD+43$^{\circ}\,3654$ is marked with a cross.
The arrow represents the velocity of the star, derived from proper motions
corrected for local motion of the surrounding ISM (see text).
Synthesized beams of $12''\times 12''$ are shown in the top right corners.}
\label{continuum}
\end{center}
\end{figure}

The hypothesis of a physical association between the star and the 
radio/IR features is supported by the very good agreement of the residual 
proper motion of the star ($15.7^\circ\pm 9.4^\circ$ east of north, see 
Comer\'on \& Pasquali 2007) and the direction from the star to the apsis 
of the bow shock ($8.8^\circ\pm 10^\circ$ east of north). The star 
velocity vector on the plane of the sky is represented in the upper
panel of Fig. 2.

We used the continuum images at 1.42 and 4.86 GHz to build a spectral 
index distribution map. We only considered input pixels with a 
signal-to-noise ratio $\geq$ 4. Besides this, the spectral 
index map was masked for a signal-to-noise ratio $\geq$ 10. Figure 3 shows 
the spectral index 
distribution and corresponding noise maps.

\begin{figure}
\begin{center}
\includegraphics[bb= 160 115 500 750,width=5cm,angle=-90]{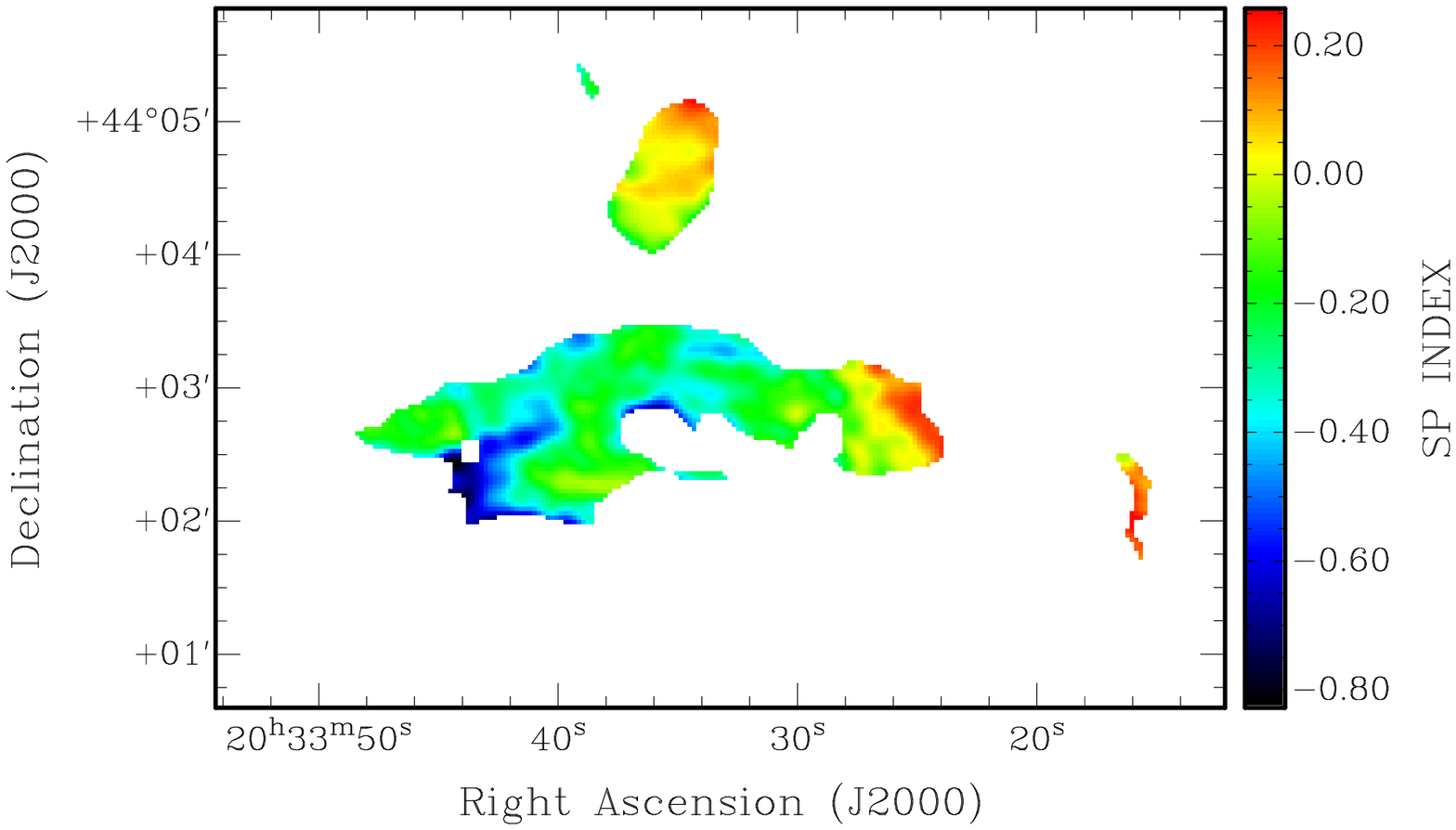}
\includegraphics[bb= 140 115 480 750,width=5cm,angle=-90]{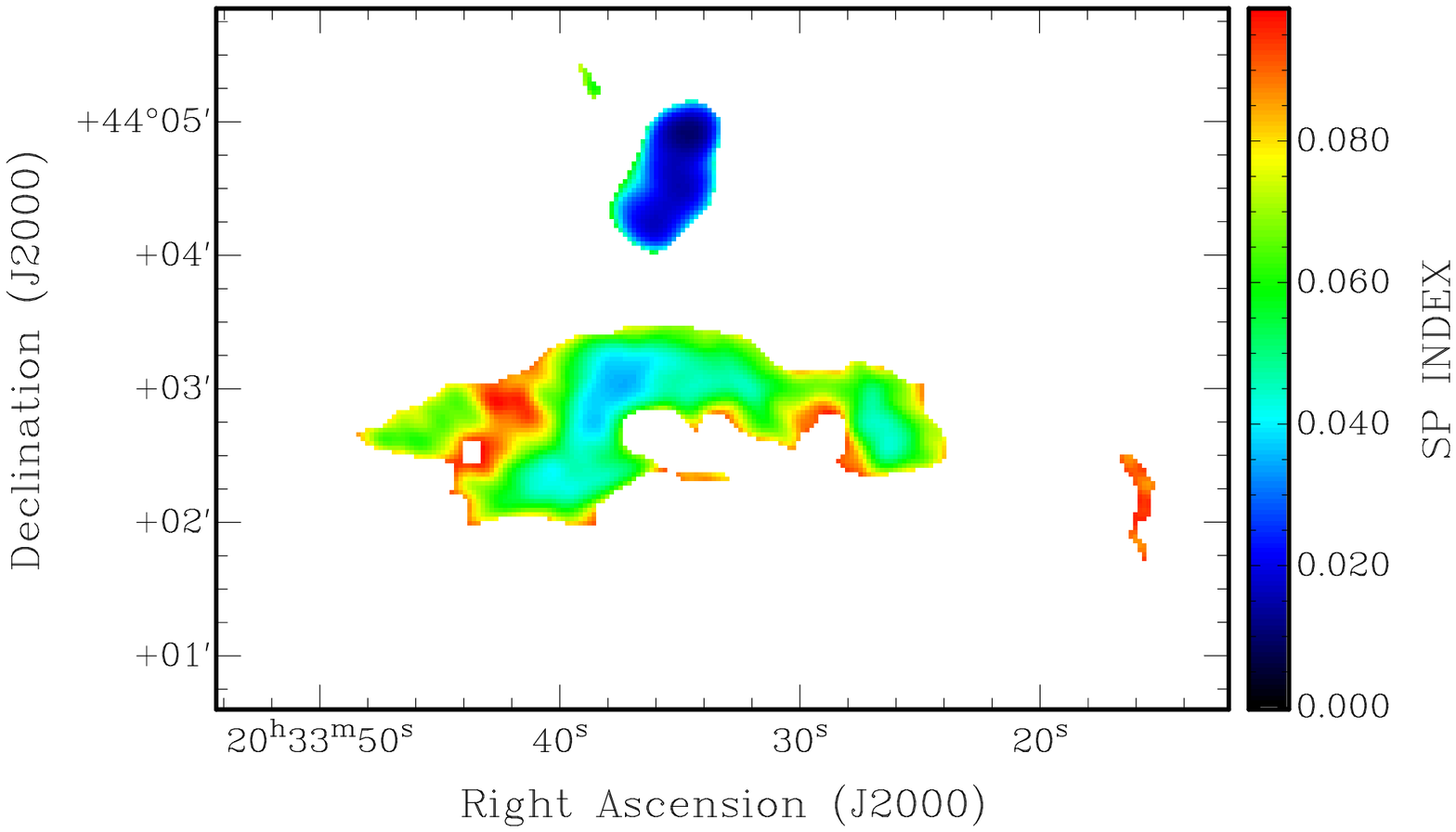}
\caption{Top: Spectral index distribution. Bottom: Spectral index error 
distribution.}
\label{spixnoise}
\end{center}
\end{figure}

\begin{figure}
\begin{center}
\includegraphics[width=6cm,angle=-90]{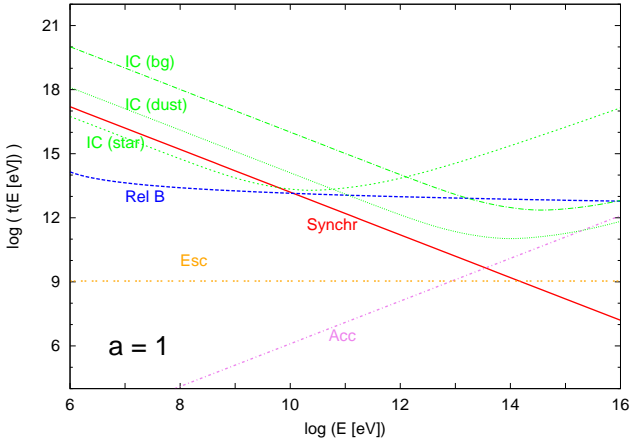}
\caption{Acceleration (`Acc'), escape (`Esc'), and cooling times for 
electrons, due to synchrotron radiation (`Synchr'), due to inverse Compton 
scattering of dust photons (`IC (dust)'), stellar photons 
(`IC (star)'), and backgroud photons (`IC (bg)'). Cooling time for relativistic 
Bremsstrahlung radiation indicated as `Rel  B'. The figure is for the case with 
equal energy density in electrons and protons ($a=1$, see 
text).}
\label{elosses}
\end{center}
\end{figure}

\begin{figure}
\begin{center} 
\includegraphics[
width=6cm,angle=-90]{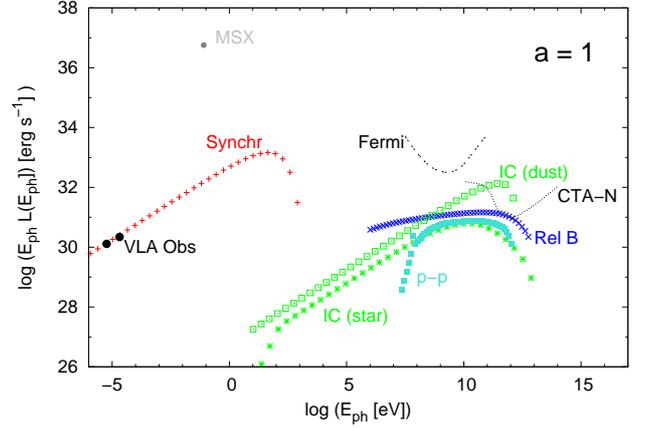}
\caption{
Spectral energy distribution for the case where the density of primary 
electrons is equal to the density of protons ($a=1$). Acronyms as in Figure
4. Measured radio fluxes from VLA observations: `VLA Obs'. The MSX 
luminosity at D band is also represented. The contribution from 
secondary pairs is 
negligible in this case, so is not shown here.}
\label{SED-1}
\end{center}
\end{figure}

\section{Nonthermal radio emission from the bow shock}

In Figure 3 we have represented the spectral index values derived from 
1.42 and 4.86 GHz data at positions where {\sl (i)} the radio continuum 
signals 
are well above the noise (4 times or more) and {\sl (ii)} the 
values of the 
spectral index errors are low ($< 10\%$). It can be appreciated from the 
figure that there is a significant fraction of the portrayed area that 
shows negative spectral indices. The $\alpha$-values range from $-$0.8 
to $+$0.3, approximately from W to E. The source ES, in 
contrast, is characterized by $\alpha \rightarrow 0$, and 
it looks unrelated to the bow-shock feature.

In what follows, we assume that the MSX-D\&E bands excess emission represents 
the projection of the bow shock in the plane of the sky. We adopt a distance 
to the bow shock of 1.4 kpc (Hanson 2003, Comer\'on \& Pasquali 2007). We 
measured a distance from the star BD$+43^{\circ}$3654 to the bow shock 
$R=5'$, or 2 pc (at 1.4 kpc). The shape of the bow shock can be 
fitted to a spherical cap of height $2'= 0.8$ pc. We  
approximated the area of 
the MSX structure on the plane of the sky by $6'$ in length times $2'$ in 
height. If the depth is equal to the length, the volume occupied by the 
bow shock is $\sim 4.6\, {\rm pc}^3$. Kobulnicky et al. (2010) 
suggest a conservative lower limit for the
particle density in the bow-shock region of 100 cm$^{-3}$. The area 
circumscribed by the projection of the cap is $\sim$1.8 pc$^2$. Most of this 
area shows a source of nonthermal radiation with averaged index 
$\alpha=-0.5$. 
Such a radiation is expected from synchrotron emission generated by 
relativistic electrons accelerated either at the forward shock in the ISM or 
in the reverse shock in the stellar wind. Since the latter is expected to 
have velocities $v_{\rm rs}\sim v_{\rm w}\approx 2300$ km s$^{-1}$, it should 
be more efficient for electron acceleration (as the efficiency 
$\eta\propto(v_{\rm rs}/c)^{2}$ for first-order Fermi-like diffusive 
acceleration, e.g. Drury 1983). The forward shock moves in a denser medium 
with lower velocities, $\sim 400$ km s$^{-1}$, but it might be effective to 
accelerate electrons as well although only up to lower energies.

To estimate the number density of relativistic particles, we 
followed the standard method by using the observed flux density and 
spectral slope, along with the hypothesis of equipartition between 
magnetic field and relativistic particles energy density (e.g. Guinzburg 
\& 
Syrovatskii 1964, Araudo et al. 2007). We considered that the 
energy density of relativistic particles makes three contributions (addends): 
\begin{eqnarray}
	u&=&\int E_{e1} n_{e1}(E_{e1}) dE_{e1}+ \nonumber \\ 
	&& +\int E_{p} n_{p}(E_{p}) dE_{p}+\int E_{e2} n_{e2}(E_{e2}) dE_{e2},
\end{eqnarray}
where $e1$, $p$, and $e2$ stand for relativistic primary electrons, protons, and 
secondary electron-positron pairs (i.e. pairs coming from charged 
pion decays), respectively, and $n$ is 
the number density. 
The relation between primary electrons and protons is $u_{p}=a u_{e1}$, with 
$a\geq0$. Three cases were considered: $a=0$ (just electrons), $a=1$ (equal 
energy density in both species), and $a=100$ (proton-dominated case, as observed 
in the galactic cosmic rays). The order of magnitude of the equipartition 
magnetic field led to $B\sim 5 \times 10^{-5}$ G (three times this value for 
the case $a=100$). The maximum value for the energy of the particles was 
determined through the balance of energy gain and losses. Different loss 
mechanisms were considered: synchrotron radiation, inverse Compton (IC) 
scattering of IR, stellar, cosmic microwave background photons, relativistic 
Bremsstrahlung, and particle escape from the radiation region due to convection 
by the stellar wind. In the case of protons, the only relevant losses are 
proton-proton ($pp$) inelastic collisions and convective escape. Diffusion is 
negligible in comparison to convection in this situation. Both primary 
electrons and protons reach energies up to $\sim 10^{13}$ eV, which is 
imposed by 
nonradiative losses, except for $a=100$, where synchrotron losses
dominate
for electrons. In Fig. \ref{elosses} we show 
the losses for electrons in the case $a=1$. Values of magnetic field
and maximum 
energies obtained for electrons and protons are given in Table 1.

%


\section{Discussion and perspectives}

The presence of highly relativistic particles in a dense medium with  
high photon density can result in the efficient generation of gamma-rays. 
Although protons can be effectively accelerated up to the highest 
energies only in the shocked wind, where the density is low, they can 
diffuse or be convected upstream up to the region with the swept material and 
densities 
of $n\sim 100$ cm$^{-3}$. The corresponding gamma-ray emissivity can be 
calculated using the delta-functional approximation (e.g. Aharonian \& 
Atoyan 2000, Kelner et al. 2006). For the case $a=1$, the total luminosity 
from $pp$ interactions is similar to what is obtained from relativistic 
Bremsstrahlung of electrons, since the cross sections are similar. In 
Figure \ref{SED-1} we show the spectral energy distribution obtained for 
the case $a=1$, with all contributions included (synchrotron 
self-Compton losses are negligible). It can be seen that in 
this case the inverse Compton up-scattering of IR photons is the major 
contribution at high energies, with a peak around 100 GeV. 
The detectability of the source by instruments like the Fermi 
$\gamma$-ray observatory LAT will depend on the actual particle density
and the contribution related to the secondary electrons at large $a$.
Detailed calculations for a set of main parameters will be given 
elsewhere.
The $pp$ contribution extends well into the TeV regime, but 
it is weaker and will be difficult to detect with ground-based Cherenkov 
telescope arrays like VERITAS or MAGIC II. In contrast, if the 
relativistic particle contain is proton-dominated ($a=100$), gamma-rays 
from neutral pion decay dominate the high-energy spectrum. 
The planned CTA North observatory might detect the source, easily yielding 
imformation on the cutoff at high energies.
Observations 
of the spectral slope in this regime can be used to identify the proton 
content through the luminosity level and the proton spectral index, 
since it is preserved in the corresponding photon index. Radio 
polarization data will provide additional information on the magnetic 
field. Observations of BD+43$^{\circ}\,3654$ with X-ray observatories 
like {\sl XMM-Newton} and {\sl Chandra} can be very useful for determining 
the cutoff of the synchrotron spectrum, which is directly related to the 
maximum energy of the electrons. This, in turn, would yield valuable 
information on the actual value of the magnetic field and the correctness 
of the equipartition hypothesis.

\begin{table}[]
\begin{center}
\caption{Magnetic field and maximum energies calculated 
for the three cases of $a$ (see text) for primary electrons and
maximum energy for protons.}\label{tabla1}
\begin{tabular}{cccc}
\hline
\hline 
$a$  & $B$ & $E_{\rm max}^{e1}$ & $E_{\rm max}^{p}$ \\  
   & [G] & [eV] & [eV]\\ 
\hline 
$0$& $4.0\times10^{-5}$  & $8\times10^{12}$ & -- \\  
$1$& $5.0\times10^{-5}$  & $9\times10^{12}$ &  $9\times10^{12}$ \\  
$100$&$1.4\times10^{-4}$ & $2\times10^{13}$  & $3\times10^{13}$ \\
\hline
\hline 
\end{tabular}
\end{center}
\end{table}


\begin{acknowledgements}
This work was supported by MinCyT - ANPCyT, project number PICT-2007-00848 / 
Pr\'estamo BID, and by CONICET, project ID 11220090100078. JM and GER acknowledge 
support by grant AYA2007-68034-C03-01 and -02 from the Spanish government and 
FEDER funds. JM and GER are also supported by Plan Andaluz de Investigaci\'on, 
Desarrollo e Innovaci\'on of Junta de Andaluc\'{\i}a as research group FQM-322 
and excellence fund FQM-5418.
\end{acknowledgements}

\end{document}